\documentclass[onecollarge]{svjour2}
\smartqed  
\usepackage{graphicx}
\usepackage{mathptmx}      
\usepackage{amsfonts}
\usepackage{amssymb}
\usepackage{bm} 
\usepackage{bbm}
\usepackage{amsmath}
\usepackage{slashed}
\journalname{Few-Body Systems (EFB22)}
\begin{document}

\title{Singularity-free two-body equation with confining interactions in momentum space
}

\titlerunning{Singularity-free two-body equation with confining interactions}        

\author{Alfred Stadler  \and
        Sofia Leit\~ao \and
       M.\ T.\  Pe\~na \and
        Elmar P.\ Biernat
}

\authorrunning{A.\ Stadler, S.\ Leit\~ao, M.\ T.\  Pe\~na, E.\ Biernat} 

\institute{Alfred Stadler \at
   Departamento de F\'isica da Universidade de \'Evora, 7000-671 \'Evora, Portugal\\
  and Centro de F\'isica Nuclear da Universidade de Lisboa, 1649-003 Lisboa, Portugal\\
  \email{stadler@uevora.pt}
           \and
           Sofia Leit\~ao, M.\ T.\  Pe\~na, and Elmar Biernat \at
             Departamento de F\'isica and CFTP, Instituto Superior T\'ecnico, Av. Rovisco Pais, 
1049-001 Lisboa, Portugal
}

\date{Received: date / Accepted: date}

\maketitle

\begin{abstract}
We are developing a covariant model for all mesons that can be described as quark-antiquark bound states in the framework of the Covariant Spectator Theory (CST) in Minkowski space. The kernel of the bound-state equation contains a relativistic generalization of a linear confining potential which is singular in momentum space and makes its numerical solution more difficult. The same type of singularity is present in the momentum-space Schr\" odinger equation, which is obtained in the nonrelativistic limit. We present an alternative, singularity-free form of the momentum-space Schr\" odinger equation which is much easier to solve numerically and which yields accurate and stable results. The same method will be applied to the numerical solution of the CST bound-state equations.
\end{abstract}

\section{Introduction}
\label{intro}
This work is part of an effort to develop a manifestly covariant model for a unified description of all mesons that can be understood as $q\bar{q}$ states bound by a confining interaction. Mesons containing only heavy quarks are essentially nonrelativistic systems, but relativity is an essential ingredient in a theory that includes light quarks. In addition, the requirements of chiral symmetry have to be respected for a realistic description of the pion.
Our theoretical framework is the Covariant Spectator Theory (CST), which is based on Relativistic Quantum Field Theory and can be viewed as a reorganization of the Bethe-Salpeter equation (for a recent brief review see Ref.~\cite{Sta11}). By incorporating the quark self-interaction through the same relativistic kernel that describes the interaction between two different quarks, we improve on previous work by Gross, Milana, and \c Savkli 
\cite{GM-qqbar,Savkli:1999me}
and make the dynamics self-consistent \cite{Biernat:2013fkaVAR1}.

\begin{figure}[tbh]
\centering
\includegraphics[width=\columnwidth]{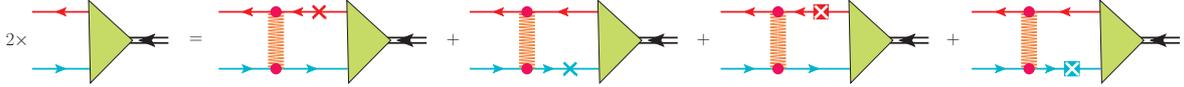}
\caption{The CST--Bethe-Salpeter vertex function calculated in terms of the four CST vertex functions. A cross on a quark line indicates that only the positive-energy pole contribution of the corresponding propagator is used in the loop integration. A cross inside a square refers to the respective negative-energy pole.}
\label{Fig:CST-BS}
\end{figure}

Figure \ref{Fig:CST-BS} shows how the $q\bar{q}$ CST--Bethe-Salpeter vertex function with both external quark momenta off mass shell is related to four distinct CST vertex functions. It corresponds to the usual Bethe-Salpeter equation, except that only propagator pole contributions are included in the loop integration. To determine the four CST vertex functions, each of the two external quark momenta is, one at a time, placed on its positive- or negative-energy mass shell. The four possible choices then yield a closed system of four equations, referred to as the ``CST four-channel equations''. It can be shown that they are charge-conjugation symmetric as required for the description of equal-mass $q\bar{q}$ systems \cite{Biernat:2013fkaVAR1}.

The CST bound-state equations are homogeneous integral equations formulated in momentum space. A relativistic generalization of a linear potential is used as confining interaction, to which constant or color-Coulomb interactions can be added. Our goal is to construct a model that provides an accurate description of both the energy spectrum and the meson structure, the latter of which is probed in elastic and transition form factors. 
However, in order to determine the solutions of the CST equations, reliable numerical methods for dealing with confining interactions in momentum space are required. 

To develop such a method we study the nonrelativistic limit in which the CST equation becomes the Schr\" odinger equation. This is very convenient because exact solutions of the Schr\" odinger
 equation with a linear potential for S-waves are known in coordinate space in terms of Airy functions. The validity of our numerical techniques can therefore be tested by comparison with the exact results.

\section{Linear confinement in momentum space}
\label{Sec:Conf}
It is {\it a priori} not obvious how best to treat the linear coordinate-space potential $W_L({\bf r})=\sigma r$ in momentum space.  One way is to start from a screened version of the form
\begin{equation}
 W_{L,\epsilon}({\bf r})
 =\frac{\sigma}{\epsilon} \left( 1-e^{-\epsilon r} \right) =W_{A,\epsilon}({\bf r})-W_{A,\epsilon}(0) \, ,
 \label{eq:WLeps}
\end{equation} 
where
\begin{equation}
 W_{A,\epsilon}({\bf r})
 = -\frac{\sigma}{\epsilon}e^{-\epsilon r} \, ,
 \label{eq:WepsA}
\end{equation} 
and $W_L({\bf r})=\lim_{\epsilon \rightarrow 0}W_{L,\epsilon}({\bf r})$.
After calculating the Fourier transform of $W_{A,\epsilon}(\mathbf{r})$,
\begin{equation}
 V_{A,\epsilon}(\mathbf{q})=\int d^{3}r\, 
W_{A,\epsilon}(\mathbf{r}) e^{i\mathbf{q}\cdot\mathbf{r}} 
= -\frac{8\pi\sigma}{(q^2+\epsilon^2)^2} \, , 
\end{equation} 
and observing that
\begin{equation}
\int  \frac{d^3 q}{(2\pi)^3}  \, V_{A,\epsilon}({\bf q}) = -\frac{\sigma}{\epsilon} \, ,
\end{equation}
one can write the screened linear potential in momentum space as
\begin{equation}
 V_{L,\epsilon}({\bf q})  = \int d^3r \,  W_{L,\epsilon}({\bf r}) e^{i{\bf q}\cdot {\bf r}}
= V_{A,\epsilon}({\bf q})
 - (2\pi)^3 \delta^{(3)}({\bf q}) \int  \frac{d^3 q'}{(2\pi)^3} V_{A,\epsilon}({\bf q}') \, .
\end{equation}
The unscreened limit becomes
\begin{equation}
V_L({\bf q}) = \lim_{\epsilon \rightarrow 0} V_{L,\epsilon}({\bf q})
= \left[ V_A({\bf q}) -(2\pi)^3\delta^{(3)}({\bf q})\int \frac{d^3 q'}{(2\pi)^3} V_A({\bf q}') \right] \, ,
\label{eq:VL}
\end{equation}
where
\begin{equation}
V_A(\mathbf{q})=
 -\frac{8\pi\sigma}{{\bf q}^4} \, .
 \label{Eq:VA}
\end{equation} 
The form (\ref{Eq:VA}) suggests that a covariant generalization of the linear potential in momentum space is obtained by the replacement ${\bf q}^2 \rightarrow -q^2$, where $q$ is the four-vector of the transferred momentum. This generalization guarantees that the correct linear potential is reproduced in the nonrelativistic limit. That it is also strictly confining when used in the kernel of the CST bound-state equation has been demonstrated in Ref.~\cite{Savkli:1999me}.

Substituting the unscreened linear confining kernel (\ref{eq:VL}) into the momentum-space Schr\" odinger equation for a system with reduced mass $m_R$ yields 

\begin{equation}
\frac{p^2}{2m_R}  \psi( {\bf p}) 
+  
\int  \frac{d^3 k}{(2\pi)^3}  \, V_A({\bf p}-{\bf k})
\left[  \psi( {\bf k})  
 - \psi( {\bf p})  \right] =
 E  \psi( {\bf p}) \, .
 \label{eq:SE}
\end{equation}
Next we perform a projection of Eq.\ (\ref{eq:SE}) onto partial wave $\ell$, which leads to
\begin{equation}
\frac{p^2}{2m_R} 
\psi_\ell(p) 
+  
\int_0^\infty  \frac{d k \, k^2}{(2\pi)^3}
\left[
V_{A,\ell}(p,k)
\psi_\ell(k) - 
V_{A,0} (p,k)
\psi_\ell(p) 
 \right]
   =
 E \psi_\ell(p)  \, .
 \label{eq:SPW}
\end{equation}
The kernel $V_A$ in partial wave $\ell$ is
\begin{equation}
V_{A,\ell}(p,k)  =
2\pi (-8\pi\sigma)
\left[ 
\frac{2 P_\ell(y)}{\left( p^2-k^2 \right)^2}
-
\frac{P'_\ell(y)}{\left( 2p k \right)^2} \ln \left( \frac{p+k}{p-k}\right)^2
+
\frac{2 w'_{\ell-1}(y)}{\left( 2p k \right)^2}
\right] \, ,
\label{eq:VAPW}
\end{equation}
where $P_\ell$ are the Legendre polynomials, and
\begin{equation}
 w_{\ell-1}(y) = \sum_{m=1}^\ell \frac{1}{m} P_{\ell-m}(y) P_{m-1}(y)  \, , \quad \mbox{with} \quad y=\frac{p^2+k^2}{2pk} \, .
\end{equation}
The kernel (\ref{eq:VAPW}) is highly singular at $k=p$. However, the subtraction term in the integrand of (\ref{eq:SPW}) reduces the singularity to one of Cauchy principal-value type. This subtraction term arises automatically with our choice (\ref{eq:WLeps}) of the screened linear potential. In other approaches with different screening, the same final result is obtained with different methods \cite{GM-qqbar,Spence:1987aa,Maung:1993aa}.

Although Eq.\  (\ref{eq:SPW}) can be solved numerically, the principal-value singularity renders its numerical solution rather cumbersome, and in some cases it is very difficult to obtain stable converged results.
We have found that this principal-value singularity can be eliminated by another subtraction, and that the resulting equation (\ref{eq:SPWNS}) is in fact much easier to solve: 
\begin{multline}
\left[
\frac{p^2}{2m_R} + \frac{\sigma \pi \ell(\ell+1)}{4p} \right]  \psi_\ell(p)
-
 \frac{2\sigma}{\pi}\int_0^\infty dk 
  \left\{ 
 \frac{1}{k^2-p^2}
 \left[  \frac{2k^2}{k^2-p^2}   \Big( P_\ell(y) \psi_\ell(k)- \psi_\ell(p) \Big)  -p \psi'_\ell(p)  \right]
\phantom{\left(\frac{p+k}{p-k} \right)^2} 
 \right.  \\
\left. 
+ \frac{w'_{\ell-1}(y)}{2p^2} \psi_\ell(k)
 -\frac{1}{4p^2} \ln \left( \frac{p+k}{p-k} \right)^2 
 \left[ P'_\ell(y) \psi_\ell(k) - \frac{\ell(\ell+1)}{2}\frac{p}{k}\psi_\ell(p) 
 \right]
 \right\}
 = E \psi_\ell(p) \, .
 \label{eq:SPWNS}
\end{multline}
The additional subtraction term is the one proportional to $\psi'_\ell(p)$. It eliminates the principal-value singularity in all partial waves. At first glance it may appear as a disadvantage to introduce a dependence on the derivative of the unknown function into the problem. However, when Eq.~(\ref{eq:SPWNS}) is solved by expanding $\psi_\ell(p)$ in a set of basis functions, calculating the derivative is actually trivial.   

The logarithmic term appears only for $\ell \ge 1$. It is also singular at $k=p$, but this singularity is integrable. Nevertheless, it is also subjected to a subtraction already well-known in the literature, which considerably improves convergence of the numerical solution of  (\ref{eq:SPWNS}). The term involving $w'_{\ell-1}(y)$ is nonzero only for $\ell \ge 2$ and does not introduce any additional singularities. Thus, Eq.~(\ref{eq:SPWNS}) is completely singularity-free.

\section{Numerical results}
\label{Sec:Num}
\begin{figure}[t]
\begin{minipage}[b]{0.49\linewidth}
\centering
\includegraphics[width=\columnwidth]{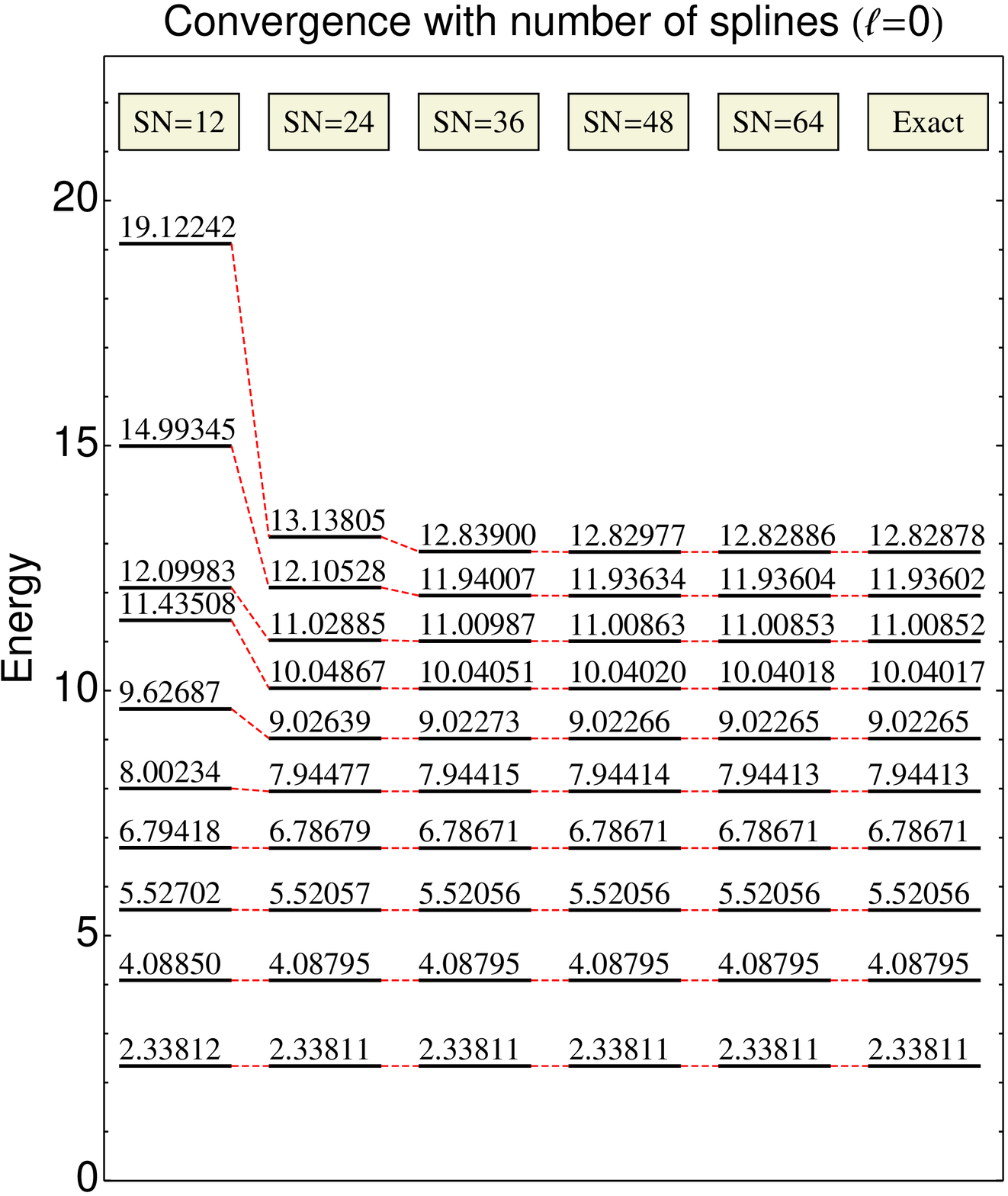}
\end{minipage}\hspace*{1mm}
\begin{minipage}[b]{0.49\linewidth}
\centering
\includegraphics[width=\columnwidth]{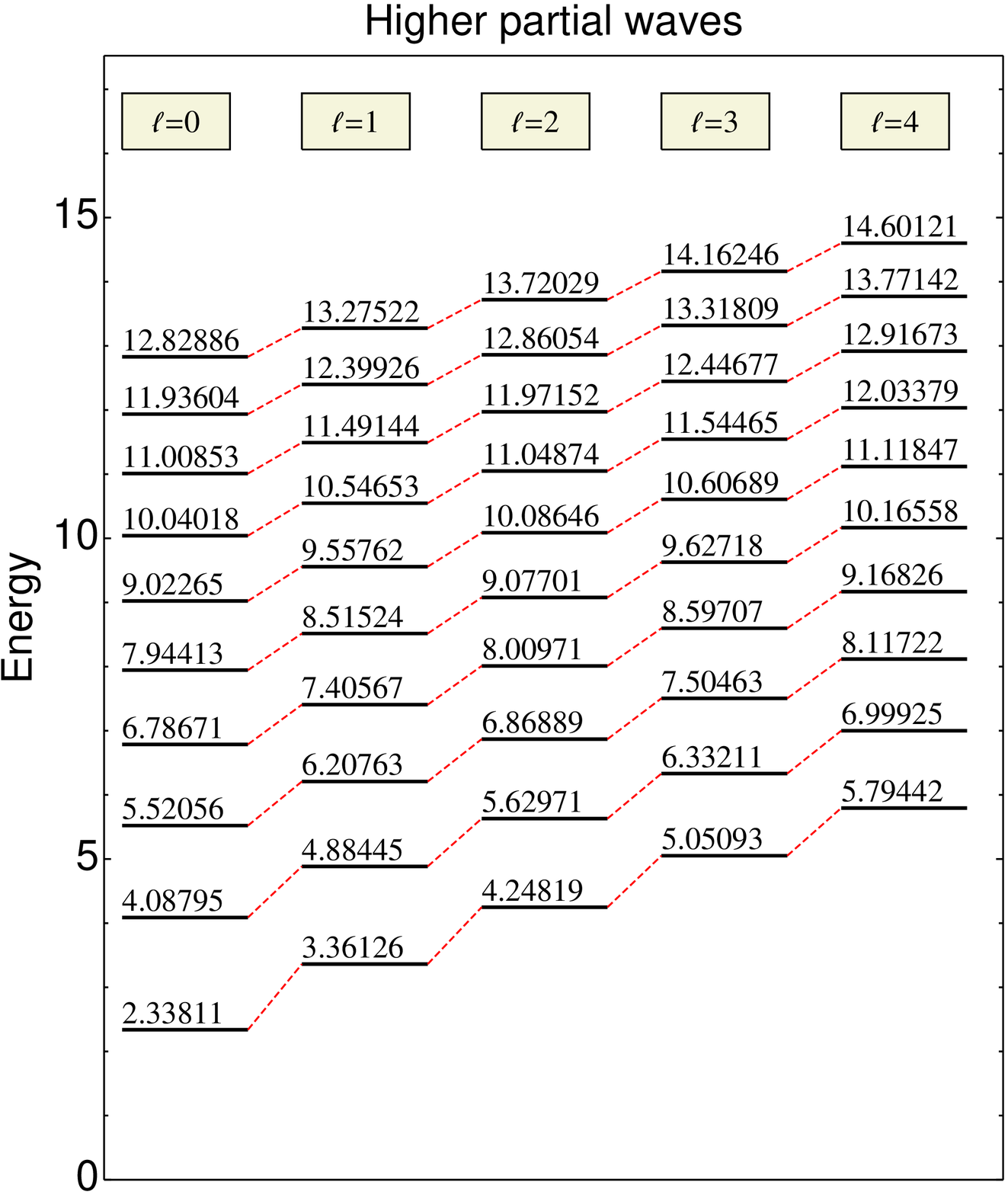}
\vspace*{-15pt}
\end{minipage}
\caption{The ten lowest bound-state energies of the linear potential $V(r)=\sigma r$, obtained as solutions of the singularity-free equation in momentum space. The left panel shows that the solutions for $S$ waves rapidly converge to the exact result with increasing number of splines, SN, in the expansion basis. The right panel shows the ten lowest energy states for the partial waves from $\ell = 0$ to $\ell=4$, using $\mbox{SN}=64$. The energies are given in units of $(\sigma^2/2 m_R)^{1/3}$, where $m_R$ is the reduced mass of the $q\bar{q}$ system.}
\label{Fig:results}
\end{figure}

In our numerical test calculations we chose to expand the wave functions $\psi_\ell(p)$ in a basis of B-splines. For the case of S-waves, the left panel of Fig.~\ref{Fig:results} shows that the energy eigenvalues converge quickly to the exact solutions as the number of splines in the basis increases. We numerically Fourier transformed the obtained S-wave eigenfunctions and found that they are also in excellent agreement with the exact solutions given in terms of Airy functions. 


The eigenstates in higher partial waves, for which exact solutions are not known, can be obtained with the same method. As an example, the lowest-lying energy eigenvalues for partial waves up to $\ell=4$ are displayed in the right panel of Fig.~\ref{Fig:results}.

Apart from serving for the purpose of numerical tests, these results are also interesting by themselves. For heavy quarkonia the nonrelativistic description is known to be a good approximation, and a comparison with experimental data becomes meaningful. The charmonium and bottomonium spectra calculated with the CST equation should be very close to the ones obtained with the Schr\" odinger equation, and thus the latter can be used as benchmark results.

The type of the singularity in the linear confining kernel of the CST equation is the same as in the Schr\"odinger equation. We are therefore confident that we have not only found an efficient numerical method to solve the Schr\" odinger equation with a linear confining potential in momentum space, but that the same method to eliminate the singularity  can also be applied to
determine the solutions of the CST bound-state equation for $q\bar{q}$ systems.

\begin{acknowledgements}
This work received financial support from Funda\c c\~ao para a Ci\^encia e a Tecnologia (FCT) under grant Nos.~PTDC/FIS/113940/2009, CFTP-FCT (PEst-OE/FIS/U/0777/2013) and POCTI/ISFL/2/275. This work was also partially supported by the European Union under the HadronPhysics3 Grant No. 283286.
\end{acknowledgements}

\bibliographystyle{spmpsci}      
\bibliography{PapersDB-v1.4}   

%
%

\end{document}